\documentclass[fleqn,usenatbib]{mnras}

\usepackage{newtxtext,newtxmath}

\usepackage[T1]{fontenc}

\DeclareRobustCommand{\VAN}[3]{#2}
\let\VANthebibliography\thebibliography
\def\thebibliography{\DeclareRobustCommand{\VAN}[3]{##3}\VANthebibliography}

\usepackage{graphicx}	
\usepackage{amsmath}	
\usepackage{mathtools}
\usepackage{gensymb}
\usepackage{CJKutf8}
\usepackage{longtable,booktabs,etoolbox,multirow,tabularx}

\def\mr{\mathrm}
\def\msunyr{{{\rm M}_\odot\rm\,yr^{-1}}}
\def\msun{M_\odot}

\title[Spectropolarimetry of the TDE AT 2019qiz]{Spectropolarimetry of the tidal disruption event AT\,2019qiz: a quasispherical reprocessing layer}

\author[Patra et al.]{
Kishore C. Patra$^{1, \dagger,}$\thanks{E-mail: kcpatra@berkeley.edu},
Wenbin Lu$^{2, 1}$, 
Thomas G. Brink$^{1}$, 
Yi Yang$^{1, \mathsection}$, 
Alexei V. Filippenko$^{1}$,  \newauthor  
Sergiy S. Vasylyev$^{1, \ddagger}$
\\
$^{1}$Department of Astronomy, University of California, Berkeley, CA 94720-3411, USA\\
$^{2}$Department of Astrophysical Sciences, Princeton University, Princeton, NJ 08544, USA\\
$^{\dagger}$Nagaraj-Noll-Otellini Graduate Fellow \\
$^{\mathsection}$Bengier-Winslow-Robertson Postdoctoral Fellow\\
$^{\ddagger}$Steven Nelson Graduate Fellow\\
}

\date{Accepted XXX. Received YYY; in original form ZZZ}

\pubyear{2022}

\begin{document}
\label{firstpage}
\pagerange{\pageref{firstpage}--\pageref{lastpage}}
\maketitle

\begin{abstract}
We present optical spectropolarimetry of the tidal disruption event (TDE) AT\,2019qiz on days $+0$ and $+29$ relative to maximum brightness. Continuum polarization, which informs the shape of the electron-scattering surface, was found to be consistent with 0 per cent at peak brightness. On day $+29$, the continuum polarization rose to $\sim 1$ per cent, making this the first reported spectropolarimetric evolution of a TDE. These findings are incompatible with a naked eccentric disc that lacks significant mass outflow. Instead, the spectropolarimetry paints a picture wherein, at maximum brightness, high-frequency emission from the accretion disc is reprocessed into the optical band by a nearly spherical, optically thick, electron-scattering photosphere located far away from the black hole. We estimate the radius of the scattering photosphere to be  $\sim 100\rm\, au$ at maximum brightness --- significantly larger than the tidal radius ($\sim 1\rm\, au$) and the thermalisation radius ($\sim 30\rm\, au$) where the optical continuum is formed. A month later, as the fallback rate drops and the scattering photosphere recedes, the continuum polarization increases, revealing a moderately aspherical interior. We also see evidence for smaller-scale density variations in the scattering photosphere, inferred from the scatter of the data in the Stokes $q-u$ plane. On day $+29$, the H$\alpha$ emission-line peak is depolarized to $\sim 0.3$ per cent (compared to $\sim 1$ per cent continuum polarization), and displays a gradual rise toward the line's redder wavelengths. This observation indicates the H$\alpha$ line formed near the electron-scattering radius.

\end{abstract}




\section{Introduction}
\label{sec:introduction}

Occasionally a star gets too close to a supermassive black hole (SMBH), at which point tidal forces on the star exceed its self gravity, tearing apart the ill-fated interloper. Tidal disruption events (TDEs) --- as these transient events are called --- occur when a star wanders within the tidal radius, $r_{\text{t}} \approx R_{\star} (M_{\text{BH}}/M_{\star})^{1/3}$, of the SMBH \citep{Hills_1975}. Owing to the large spread in specific orbital energy acquired by the post-disruption stellar debris, roughly half of the debris is bound in a range of highly-eccentric orbits around the SMBH, whereas the rest goes unbound \citep{Rees_1988}. The bound material is expected to fall toward the SMBH, forming an accretion disk. The gravitational potential energy lost by the infalling matter is converted into electromagnetic radiation, resulting in a luminous flare that enables the detection of TDEs.

All TDEs were originally expected to be extreme-ultraviolet (UV) and X-ray emitters, as is the case for active galactic nuclei. However, many of them were found to show bright near-UV/optical emission with a luminosity much higher than (or at least comparable to) that in the X-ray band \citep{gezari08_GALEX_TDEs, gezari12_PS10jh, holoien16_as14li, Auchettl_etal_2018, hinkle21_as19dj, vanvelzen21_TDE_tdes}, igniting a long-standing controversy on the origin of the bright optical emission and on the apparent absence of X-rays in these TDEs\footnote{In this paper, we focus on TDEs that are selected by UV/optical transient surveys, as they are more accessible to optical spectroscopy. Another class of TDEs selected by X-ray transient surveys indeed emit the majority of their luminosity in the extreme-UV and soft X-ray bands with subdominant optical emission, and their rate is likely comparable to the optically-selected ones \citep{komossa99_ROSAT_TDEs, esquej08_XMM-Newton_TDEs, sazonov21_eROSITA_TDEs}. }. Two classes of solutions have been proposed: (i) the accretion disc does produce ample X-ray emission, but it gets reprocessed by an optically-thick gas layer located at a distance much larger than the tidal radius (e.g., \citealp{Strubbe_etal_2009, Guillochon_etal_2014, Roth_etal_2016, metzger16_reprocessing, Dai_etal_2018, Lu_Bonnerot_2020}), and (ii) instead of accretion, TDEs are powered by shocks resulting from collision of streams in the outer regions of a highly-eccentric disc (e.g., \citealp{Piran_etal_2015, Shiokawa_etal_2015}). 

Indirect evidence is beginning to mount in favor of solution (i) in the form that accretion discs are created promptly even in X-ray-lacking TDEs, with double-peaked Balmer lines \citep{Hung_etal_2020, Short_etal_2020} and Bowen fluorescence from excitation by high-energy photons \citep{Leloudas_etal_2019, Blagorodnova_etal_2019} being a few examples.

Currently, the details of how such a reprocessing gas layer comes into existence remain unclear. Proposed models include wind or outflows from the TDE \citep{Lodato_etal_2011, Miller_etal_2015_tde, Jiang_etal_2016, Dai_etal_2018, bonnerot21_first_light} or the presence of a radiation-pressure-supported envelope originating from the bound debris \citep{Loeb_Ulmer_1997, coughlin14_ZEBRA}.   

A unique and previously unseen perspective on this debate is offered by spectropolarimetry --- a technique that measures polarization \footnote{In this paper the word “polarization” refers only to linear polarization. We expect that photons emerging from the electron-scattering photosphere have no circular polarization. This is because thermal photons have random phases before they are scattered by electrons, and the scattering process itself does not generate circular polarization.} as a function of wavelength.  Light from a TDE is polarized by electron
(Thomson) scattering of photons, and the final scatter before escape
determines the photon’s polarization state. For a spatially unresolved source, such as a TDE, the total polarization will be the integration of the photons' electric vectors projected on the plane of the sky. If the sky-plane-projected photosphere is circularly symmetric, the electric vectors cancel out, resulting in zero polarization. Conversely, if the projected photosphere deviates from circular symmetry, the electric vectors undergo incomplete cancellation, resulting in a net nonzero polarization (see \citealt{Wang_Wheeler_2008} for a review). In the absence of intrinsically polarized emission from a nonthermal source (e.g., a relativistic jet), the measurement of polarization (or the lack thereof) thus informs the geometry of the electron-scattering surface. The subject of this paper is the spectropolarimetric study of the TDE AT\,2019qiz and the resulting constraints on its reprocessing layer.  

AT\,2019qiz [$\alpha$(J2000) $= 04^{\rm hr}46^{\rm m}37.88^{\rm s}$, $\delta$(J2000) = $-10^\circ13'34.90''$] was discovered on 2019 Sep. 19 \citep[UT dates are used throughout this paper;][]{Forster_2019} in the centre of the galaxy 2MASXJ04463790-1013349. A spectrum obtained on 2019 Sep. 25 by \citet{Siebert_etal_2019} showed broad He\,II and H\,I Balmer lines superposed on a blue continuum, classifying AT\,2019qiz as a TDE. In the scheme of \citet{vanvelzen21_TDE_tdes}, AT\,2019qiz falls under the TDE-Bowen class. UV/optical photometry and spectroscopy of AT\,2019qiz were analysed by \citet{Nicholl_etal_2020} and \citet{Hung_etal_2021} independently; we reference their results wherever appropriate in this paper. The TDE light curve and velocity dispersion of the host galaxy suggest that a $\sim 1$\,M$_{\odot}$ star was disrupted by a central black hole of mass $\sim 10^{6}$\,M$_{\odot}$. The peak luminosity ($\sim 10^{43}$\,erg\,s$^{-1}$) and the total integrated emitted energy ($\sim 10^{50} ~ \text{erg}$) are relatively low compared to other TDEs \citep{Nicholl_etal_2020}, which may be due to partial disruption of the star. The X-ray luminosity from this TDE was found to be 2-3 orders of magnitude lower than in the UV/optical, making it a ``normal'' optically-selected nonrelativistic TDE \citep{Nicholl_etal_2020}. The host of AT\,2019qiz is a face-on barred spiral galaxy (Hubble type SBb). A redshift of $z=0.01513$ taken from the NASA Extragalactic Database was adopted in this study. The calculated luminosity distance of 65.6\,Mpc (assuming a flat $\Lambda$CDM cosmological model with H$_{0} = 70$\,km\,s\textsuperscript{$-1$}\,Mpc\textsuperscript{$-1$} and $\Omega_{\Lambda} = 0.7$) makes AT\,2019qiz the nearest TDE discovered to date. 

The Kast spectropolarimeter on the Shane 3\,m telescope at Lick Observatory \citep{Miller_etal_1988} has a limit of $\sim 17$\,mag. Thus, most TDEs are not bright enough for a meaningful spectropolarimetric measurement by this instrument. Fortuitously, AT\,2019qiz was sufficiently close and luminous to reach a peak apparent brightness of $\sim 16$\,mag in the Zwicky Transient Facility (ZTF) $r$ band \citep{Nicholl_etal_2020}, offering a ripe opportunity for the first spectropolarimetric follow-up observations of a TDE with the Kast spectropolarimeter. In this work, we present two epochs of spectropolarimetry of AT\,2019qiz separated by a month. We describe our observations in Section~\ref{sec:observations} and present our results in Section~\ref{sec:results}. Interpretation of the data is discussed in Section~\ref{sec:interpretation} followed by a concluding summary in Section~\ref{sec:conclusion}.

\section{Observations}
\label{sec:observations}

Observations and data reduction were
carried out following the procedure laid out by \citet{Patra_etal_2022_19ein}. Below we highlight the important details. 

Two epochs of spectropolarimetry of AT\,2019qiz were obtained on 2019-Oct-08 and 2019-Nov-06 using the
polarimetry mode of the Kast Double Spectrograph.
On each night, exposures of 1050\,s each were carried out at retarder-plate angles of $0^{\circ}, 45^{\circ}, 22.5^{\circ}$, and $67.5^{\circ}$ to calculate the Stokes $q$ and $u$ parameters. All observations were carried out at low airmass ($\lesssim 1.5$), which allowed us to align the slit to a position angle of $180^{\circ}$ (north-south direction) on both nights. (Kast lacks an atmospheric dispersion compensator, so it is important to observe at low airmass or align the slit along the parallactic angle; see \citealt{Filippenko_1982}.) We obtained three sets of polarimetry exposures of AT\,2019qiz on each night. This allowed us to achieve a higher signal-to-noise ratio (S/N) by median combining the repeated measurements, and to compare the measured Stokes parameters from different exposure sets against each other to check for consistency.

We observed the unpolarized standard star HD\,12021 on both nights. The average Stokes $q$ and $u$ were measured to be $<0.05$ per cent for HD\,12021, thus confirming low instrumental polarization. The same low-polarization standard star was used in the polarizance test, where we determined the polarimetric response of the Kast spectropolarimeter to be $>99.5$ per cent over the wavelength range 4600--9000\,\AA. Additionally, we observed two high-polarization standard stars (HD\,25443 and HD\,245310) to examine the instrument's accuracy, finding the measured fractional polarization and its position angle consistent with references within 0.1 per cent and $3^{\circ}$, respectively \citep{Schmidt_etal_1992, Wolff_etal_1996}.

\section{The measured polarization}
\label{sec:results}

\begin{figure*}
	\includegraphics[width=0.9\linewidth]{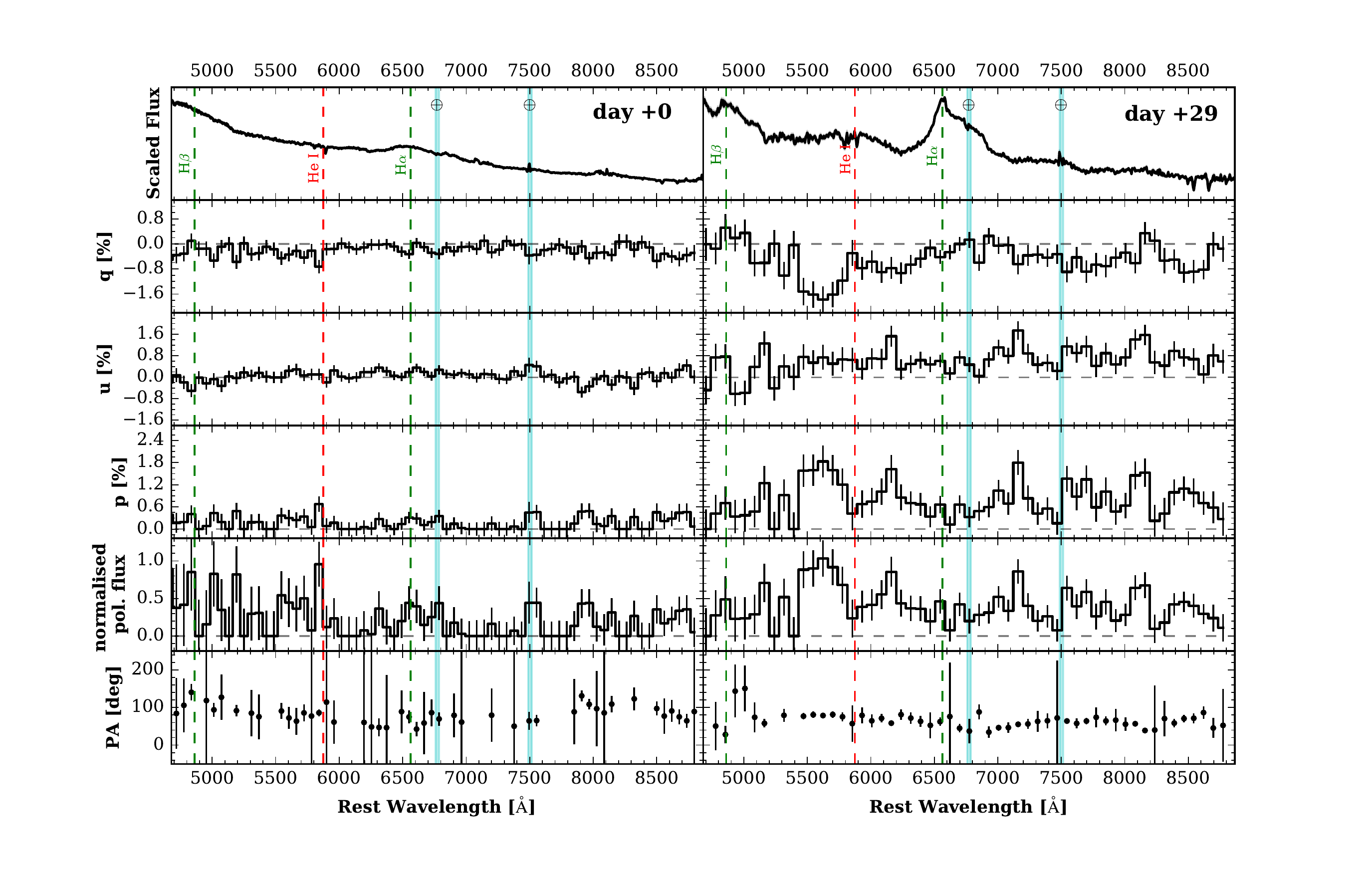}
	\caption{Spectropolarimetry of AT\,2019qiz on days $+0$ and $+29$ relative to the peak brightness on MJD~58764 \citep{Nicholl_etal_2020}. Cyan vertical bands mark the spectral regions corrected for major telluric lines. The error bars indicate 1$\sigma$ uncertainty. With the exception of the flux spectrum, we use 60\,\AA\ binning on day $+0$ and 80\,\AA\ binning on day $+29$ for clarity of presentation. Note that the two flux spectra shown here were arbitrarily scaled for better visibility. In reality, the TDE was fainter by roughly 1 magnitude on day $+29$ ($\sim 17$ mag) compared to day $+0$ ($\sim 16$ mag). $PA$ is undefined when $p$ is 0; we do not show those points in the $PA$ panel.}
	\label{fig:spol}
\end{figure*}

The intensity-normalised Stokes parameters ($q = Q/I$ and $u = U/I$, where $Q$ and $U$ are the differences in flux with electric field oscillating in two perpendicular directions, and $I$ is the total flux) were used to calculate the fractional polarization as
\begin{equation}
p_{\text{obs}} = \sqrt{q^{2} + u^{2}},  
\end{equation} 
and the polarization position angle ($PA$) as
\begin{equation}
PA_{\text{obs}} = \frac{1}{2} \arctan \left( {\frac{u}{q}} \right).
\end{equation}
The polarization position angles in this work conform to the International Astronomical Union (IAU) definition in which north = $0^\circ$ and east = $90^\circ$.
\noindent The quantity $p_{\text{obs}}$ is positive definite and thus biased toward higher polarization, especially in the low-S/N regime. Following \citet{Wang_etal_1997}, we calculated the debiased polarization as
\begin{equation}
p = \left ( p_{\text{obs}} - \frac{\sigma^{2}_{p}}{p_{\text{obs}}} \right ) \times h(p_{\text{obs}} - \sigma_{p})  ~ \mbox{and} ~~ PA = PA_{\text{obs}},
\end{equation}
where $\sigma_{p}$ denotes the 1$\sigma$ uncertainty in $p$ and $h$ is the Heaviside step function. We note that this debiasing procedure may overcorrect polarization in certain bins \citep{Montier_etal_2015}; however, we determined that using a more robust debiasing technique does not improve the result significantly. See \citet{Patra_etal_2022_19ein} for more details behind the calculation of polarization. In Figure \ref{fig:spol} we show the flux and the polarization spectra measured on days $+0$ and $+29$.

One potential complication in the analysis of the polarization of AT\,2019qiz is interstellar polarization (ISP). Nonspherical paramagnetic dust grains that are comparable in size to the wavelength of the light will preferentially extinguish photons whose polarization aligns with the long axis of the dust grains. When a large number of such dust grains --- along the sight line of the TDE --- are aligned similarly under interstellar magnetic fields, they can bias the observed polarization by inducing additional polarization. While it is generally difficult to estimate how much of the measured polarization is due to the interstellar medium, we have reasons to believe that the ISP along the direction of AT\,2019qiz is low. 

Firstly, Galactic ISP can be estimated as $9\times E(B-V)$ per cent \citep{Serkowski_etal_1975}. Considering the Milky Way colour excess $E(B-V)_{\text{MW}} = 0.09$\,mag along the line of sight of AT\,2019qiz, we find that ISP\textsubscript{MW} $\approx 0.8$ per cent \citep{Schlafly_etal_2011}. To obtain a better constraint on the Galactic ISP, we measured the polarization of an ISP ``probe star''\footnote{Gaia DR2\,3184737873690839296}, which is an intrinsically unpolarized star \footnote{Spectral type A5-F5} within $1^{\circ}$ of the line of sight of AT\,2019qiz, and distant enough to probe at least 150\,pc of the $\sim 300$\,pc scale height of the Galactic interstellar medium. (Beyond $\sim 150$\,pc, the density drops so rapidly that obtaining spectropolarimetry of a more-distant star does not improve the result much.) The measured Stokes $q$ and $u$ for the probe star were determined to be $\sim 0.1$ per cent, thus confirming low contribution from Galactic reddening. Unfortunately, the probe star does not provide any information about reddening from the host galaxy. Instead, from the host-galaxy spectral energy distribution modeling, \citet{Nicholl_etal_2020} determined a small additional extinction of $E(B-V)_{\text{host}} \approx 0.05$\,mag from the host. In combination with the Galactic ISP from the probe star, we constrain the total possible ISP to $< 0.5$ per cent along the sight line of AT\,2019qiz. This upper limit is consistent with the depolarization down to $\sim 0.3$ per cent of the H$\alpha$ and He\,I emission features. The argument here is that if these emission lines are intrinsically unpolarized, any measured polarization at those wavelengths must be due to ISP. Finally, we see no wavelength-dependent polarization on day $+0$, indicating that any dust-related polarization is likely small. Since galactic (Milky Way and host) ISP should remain constant over the lifetime of a TDE, any subsequent increase in polarization at a later time must be intrinsic to the TDE. Thus, all things considered, we deemed the ISP along the line of sight of AT\,2019qiz to be low. Justifiably, in the absence of a stronger constraint on the ISP, we did no further ISP correction.  

The continuum polarization of AT\,2019qiz at the two epochs was estimated based on the Stokes parameters over the wavelength range 7100--8200\,\AA. Admittedly, the choice of this wavelength range is arbitrary. Nonetheless, our choice was based on the fact that this region (i) is devoid of strong absorption or emission features, (ii) is sufficiently large to average over random statistical fluctuations in $q$ and $u$, and (iii) avoids the extreme ends of the spectra where the S/N is lower. Stokes $q$ and $u$ were averaged over the aforementioned wavelength range and weighted by the inverse-squared 1$\sigma$ uncertainties. The fractional continuum polarization and the associated position angle were then calculated based on the weighted mean Stokes parameters.

At its maximum brightness, the continuum polarization of AT\,2019qiz was $p_{\text{d+0}} = 0.16~\pm~ 0.05 ~(\text{stat}) \pm 0.10 ~(\text{sys})$ per cent. The systematic uncertainty is an estimation of the limitation of the instrument, which was determined from the polarimetry of bright polarization standard stars. Low polarization on day $+0$ indicates that the sky-projected electron-scattering surface is almost circularly symmetric. The polarization position angle was $PA = 85^{\circ} \pm 23$. We note that since $PA$ displays mostly random values in the low-polarization regime, this $PA$ measurement is not particularly meaningful. A month later, on day $+29$, the continuum polarization increased to $p_{\text{d+29}} = 0.93~\pm~ 0.09 ~(\text{stat}) \pm 0.10 ~(\text{sys})$ per cent. A more secure measurement of position angle was made with $PA = 59^{\circ} \pm 7$. Note that the polarization below 5500\,\AA\ is less reliable owing to comparatively poorer S/N than in the rest of the spectrum. On day$+29$, the peak of the H$\alpha$ emission line shows depolarization (compared to the continuum) down to $p_{\text{H}\alpha} = 0.3 \pm 0.1$ per cent in the 200\,\AA\ around the 6563\,\AA\ line.

\section{Interpretation}
\label{sec:interpretation}

In scenarios where optical emission of a TDE is powered by a naked eccentric disc lacking significant mass outflow, a high level of polarization can be expected owing to imperfect cancellation of electric vectors across the sky-projected surface. However, the low continuum polarization observed on day $+0$ implies that the sky-projected electron-scattering photosphere was close to being circularly symmetric. In three dimensions, it is reasonable to assume that the circular symmetry arises from the projection of a spherical electron-scattering photosphere. The continuum polarization on day $+29$ is no more than $\sim 1$ per cent, which translates to only a moderate amount of asphericity (aspect ratio $\sim 0.8$ for a Thomson scattering optical depth $\tau=1$ and a radial density profile approximated by a steady mass-loss wind; \citealp{Hoeflich_1991}). These results are incompatible with models that require a highly eccentric disc as the dominant source of optical emission. Additionally, if a disc with high initial eccentricity circularises over time, it should display high polarization at early phases followed by a gradual decrease later. The observed polarization, however, follows the opposite trend.

One might argue that a circular accretion disc is also consistent with the observed spectropolarimetry. However, such a disc requires the less-likely special condition in which it is almost face-on to the observer at maximum brightness. Weak X-ray emission observed from AT\,2019qiz (compared to UV/optical) is in tension with the face-on disc \citep{Nicholl_etal_2020}. This is because the emission of X-ray flux is viewing-angle dependent, with low X-ray emission expected from edge-on accretion discs and vice versa \citep{Dai_etal_2018}. Furthermore, the increase of polarization on day $+29$ precludes the face-on-disc scenario, unless the inclination of the disc changes dramatically over a month. The data thus disfavor a naked accretion disc --- either eccentric or circular. Instead, a quasispherical reprocessing layer originating from mass outflow is consistent with the observations, and we can place interesting constraints on the said layer.

\subsection{The size of the reprocessing layer}

The radius of the electron-scattering photosphere can be estimated from the mass outflowing rate given by $\dot{M} = 4 \pi r^{2} \rho v$, where $\rho$ is the density and $v$ is the outflow velocity. The size of the electron-scattering photosphere ($r_{\text{s}}$) is determined by the scattering optical depth $\tau_{\text{s}} = \rho \kappa_{\text{s}} r_{\text{s}}=1$, where $\kappa_{\text{s}} \approx 0.34\rm\,cm^2\,g^{-1}$ is the Thomson opacity for solar metallicity. Thus, we obtain
\begin{equation}\label{rs_v}
r_{\text{s}} = \frac{\dot{M} \kappa_{\text{s}}}{4\pi v} \approx 1.7\times10^{15}\mr{\,cm}\, {\dot{M}\over \msunyr} {10^9\mr{\,cm\,s^{-1}}\over v}.
\end{equation}
Although the detailed physical processes driving the outflow are still uncertain, we can infer the outflow velocity $v \approx 10^9\rm\,cm\,s^{-1}$ from the observed emission-line widths \citep[e.g.,][]{Hung_etal_2020}, and we expect the mass-outflow rate to be comparable to the mass-fallback rate. This is because (i) the fallback gas is only marginally bound (compared to the binding energy near the tidal radius), and (ii) the heating rate owing to shocks and viscous accretion is super-Eddington. The peak fallback rate can be estimated as $\dot{M}_{\rm fb, max} \approx M_*/3P_{\rm min} \approx 3\,\msunyr\, (M_{\rm h}/10^6\,\msun)^{-1/2} (M_*/\msun)^{1/2}$, where $M_{\rm h}$ is the black hole mass, $P_{\rm min} \approx 41\,\mr{\,d}\, (M_{\rm h}/10^6\,\msun)^{1/2} (M_*/\msun)^{1/2}$ is the minimum orbital period of the fallback material, and we have taken the stellar radius to be $R_* \approx R_\odot (M_*/\msun)$ for a main-sequence star of mass $M_*$. The mass of the central black hole of the host galaxy of AT\,2019qiz was estimated to be $\sim 10^{6}$\,M$_{\odot}$ \citep{Nicholl_etal_2020, Hung_etal_2021}. Based on these arguments, we conclude that the scattering photospheric radius of the outflow is of order $100\rm\,au$ near peak brightness (at time $t \approx P_{\rm min}$ since the disruption). At later time $t\gtrsim P_{\rm min}$, the fallback rate drops as $\dot{M}_{\rm fb}\propto t^{-5/3}$, which causes $r_{\rm s}$ to shrink.

Let us now also determine the thermalisation radius ($r_{\text{th}}$), where the optical continuum is formed. The last absorption surface is located where the effective optical depth $\sqrt{\tau_{\text{a}} \tau_{\text{s}}} \approx 1$ \citep{Rybicki_lightman_1986}, where $\tau_{\text{a}}$ and $\tau_{\text{s}}$ are the absorption and scattering optical depths, respectively.

Assuming a constant-velocity mass outflow with $\rho \propto r^{-2}$, we can write the thermalisation radius as  
\begin{equation}\label{rth_to_kappas}
r_{\text{th}} \approx \sqrt{\kappa_{\text{a}}/\kappa_{\text{s}}} ~ r_{\text{s}}.
\end{equation}
The problem here is that the absorption opacity, $\kappa_{\text{a}}$, is generally difficult to estimate from first principles (non-local thermodynamic equilibrium radiative transfer). To make progress, we can use the fact that above $r_{\text{th}}$, instead of streaming freely, photons will diffusively advance outward owing to electron scattering. Using the diffusive flux, $F_{\text{diff}} \approx U_{\text{rad}} c/\tau_{\text{s}}(r_{\text{th}})$, where $U_{\text{rad}}$ is the radiation energy density and $\tau_{\text{s}}(r_{\text{th}})=r_{\rm s}/r_{\rm th}$ is the scattering optical depth at $r_{\rm th}$, we can write the bolometric luminosity as $L = 4 \pi r_{\text{th}}^{2} F_{\text{diff}}$. Rearranging for the thermalisation radius, and using $U_{\text{rad}} = a T^{4}$, we find
\begin{equation}
r_{\text{th}} = \left(\frac{r_{\rm s} L}{4 \pi c a T^{4}} \right)^{1/3},
\end{equation} 
where $a$ is the radiation constant and $T$ is the observed color temperature. At the peak brightness of AT\,2019qiz, $L$ and $T$ were found to be about $10^{43.7}\,\text{erg\,s}^{-1}$ and $2 \times 10^{4}$\,K, respectively \citep{Nicholl_etal_2020, Hung_etal_2021}, giving $r_{\text{th}} \approx 5.7 \times 10^{14}\mr{\,cm}\,(\dot{M}_{+0}/\msunyr)^{1/3} (v/10^9\mr{\,cm\,s^{-1}})^{-1/3}$, where $\dot{M}_{+0}$ is the outflow rate on day $+0$. This shows that the optical continuum forms well below the electron-scattering surface. From here, using Equation \ref{rth_to_kappas}, we can also estimate the absorption opacity $\kappa_{\text{a}} \approx 0.1\kappa_{\rm s}$ for our fiducial parameters.

On day $+29$, the mass-outflow rate has dropped by a factor of a few (depending on $P_{\rm min}$) from the peak value, and hence the scattering photosphere would shrink significantly by the same factor. At this time, $L \approx 10^{43}\,\text{erg\,s}^{-1}$ and $T \approx 1.5 \times 10^{4}\rm\,K$, giving us $r_{\rm th} \approx 3.9\times10^{14}\mr{\,cm}\,(\dot{M}_{+29}/0.5\,\msunyr)^{1/3} (v/10^9\mr{\,cm\,s^{-1}})^{-1/3}$, which is only slightly smaller than the $r_{\text{th}}$ at maximum brightness. Thus, over time, the electron-scattering photosphere gets closer to the thermalisation radius, exposing more of the internal asymmetries that manifest as a higher-polarization signal. The recession of the photosphere also harmonises with the independent observation that X-ray emission from TDE\,2019qiz, although low initially, rose slowly and reached a peak $\sim 25$\,days after maximum brightness \citep{Nicholl_etal_2020}.

The observed rise in the level of polarization to $\sim 1$ per cent on day $+29$ can be explained by the presence of an asymmetric interior scattering photosphere, the projection in the sky of which has an aspect ratio $e \approx 0.8$ \citep{Hoeflich_1991} and is pointed $\sim 120^{\circ}$ counterclockwise from the north--south axis in the sky. This is a simplified but illustrative interpretation of the polarization signal; the electron-scattering photosphere may have more irregularities compared to an ellipsoid. The inner asphericity could arise from an asymmetric outflow in the inner regions. The implication here is that when the scattering photosphere is much larger than the thermalisation radius (as is the case at maximum brightness), the scattering surface will appear quite spherical even with an asymmetric outflow velocity field underneath. This explains why low polarization is measured near maximum brightness, but as the scattering photosphere recedes, the asymmetry of the inner regions becomes more prominent, consequently increasing the polarization.

One notable caveat to this interpretation is that the TDE spectra have some contamination from the host-galaxy stellar light. On day $+29$, as the TDE becomes fainter, the stellar light comprises a larger fraction of the observed light; note, in particular, the Ca~II~$\lambda\lambda$8498, 8542, 8662 absorption. Thus, if the stellar light were \textit{somehow} polarized, that could explain the observed rise in polarization on day $+29$. However, polarized starlight is not expected to exhibit features like those seen in the polarized-flux spectrum. Moreover, a simple argument rules out this possibility. Let us consider an upper limit of 10 per cent contamination from stellar light (the actual contamination level is probably closer to 1 per cent). If the $\sim 1$ per cent observed polarization were to come from stellar light alone, it requires the stellar light to be polarized at the 10 per cent level, which is unlikely. Any smaller amount of contamination requires the stellar light to be polarized at even higher levels.  Hence, it is reasonable to conclude that the rise in polarization on day $+29$ was not due to contamination by polarized stellar light.

We emphasise that we remain agnostic about the exact mechanism by which the outflow originates, ultimately creating the reprocessing layer. With the data at hand, we can only conclude that any model with strong mass outflow is consistent with the observations. Detailed theoretical modeling and more polarimetric TDE studies will be required to discriminate between the existing models.

\subsection{Substructures in the electron-scattering photosphere}

\begin{figure*}
	\includegraphics[width=0.8\linewidth]{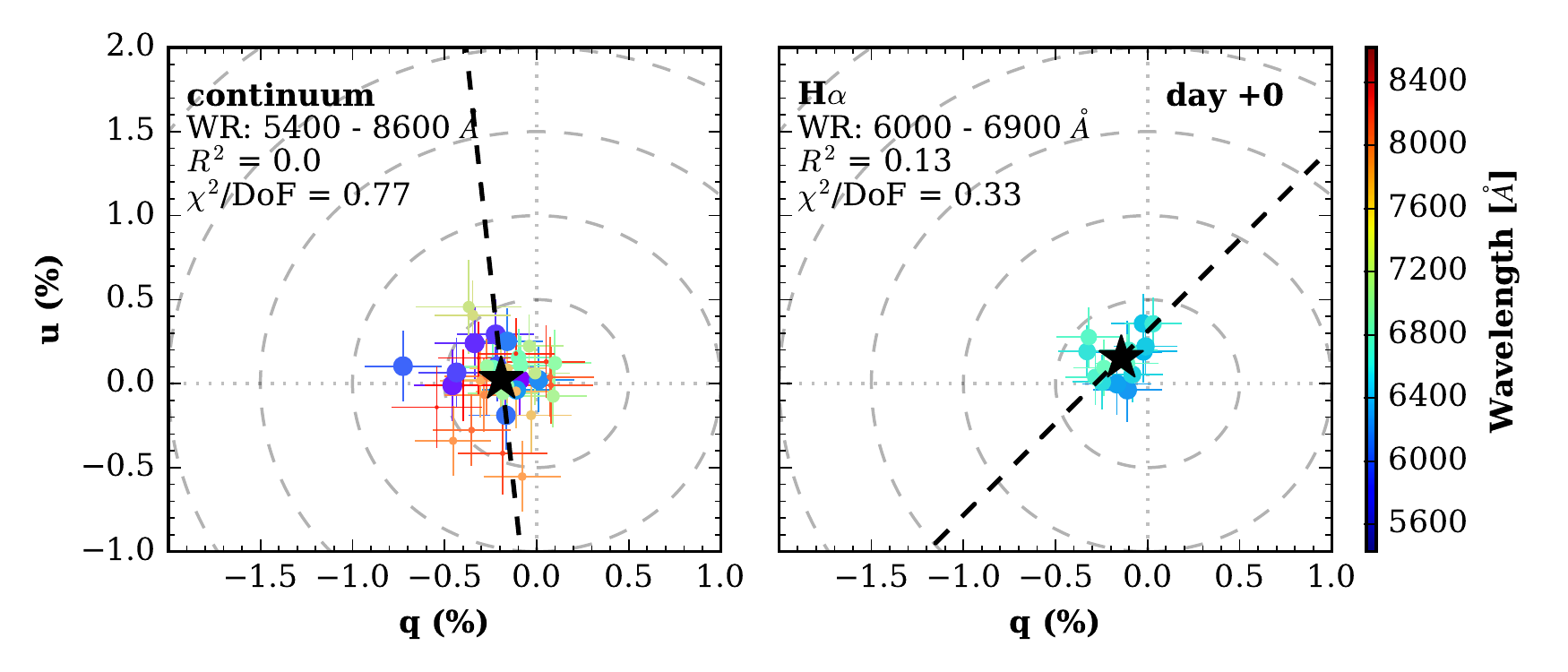}
	\caption{Polarization of AT\,2019qiz in the $q-u$ plane on day $+0$. The left-side panel shows the polarization in the wavelength range 5400--8600\,\AA, coloured by wavelength. Additionally, the size of the data points reflects their relative wavelengths in a decreasing order (i.e., larger points are bluer). The right-side panel only shows the polarization of the H$\alpha$ line in the wavelength range 6000--6900\,\AA. The black dashed lines are the best-fitting dominant axes. Constant-polarization contours in 0.5 per cent increments are shown as dashed circles. The black star in each panel represents the error-weighted mean Stokes $q$ and $u$. The coefficient of determination ($R^{2}$) and the $\chi^{2}/\text{DoF}$ are also provided.}
	\label{fig:adx_qu}
\end{figure*}

\begin{figure*}
	\includegraphics[width=0.8\linewidth]{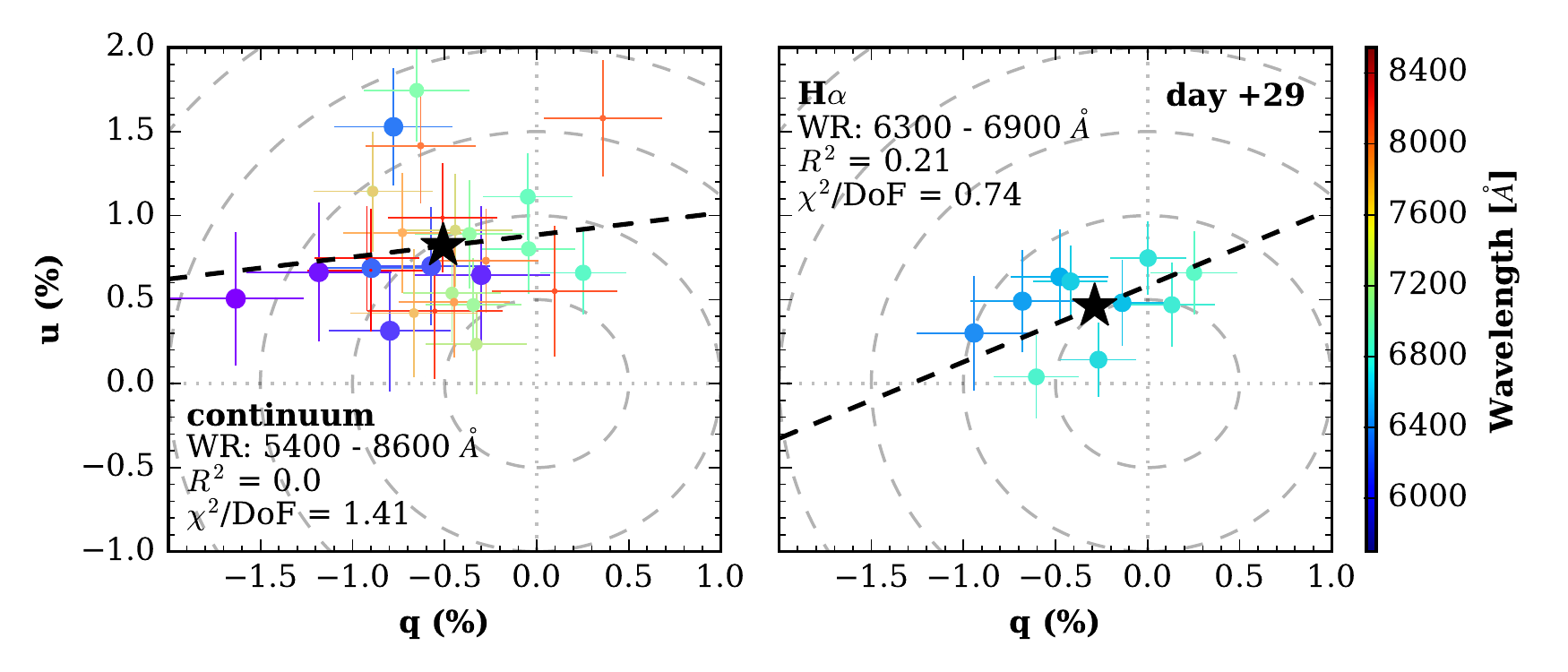}
	\caption{Same as Figure \ref{fig:adx_qu} but for day $+29$. The H$\alpha$ wavelength range is 6300--6900\,\AA. The abscissa and the ordinate are on the same scale as Figure \ref{fig:adx_qu} for ease of comparison. }
	\label{fig:aea_qu}
\end{figure*}

\begin{figure}
	\includegraphics[width=0.9\linewidth]{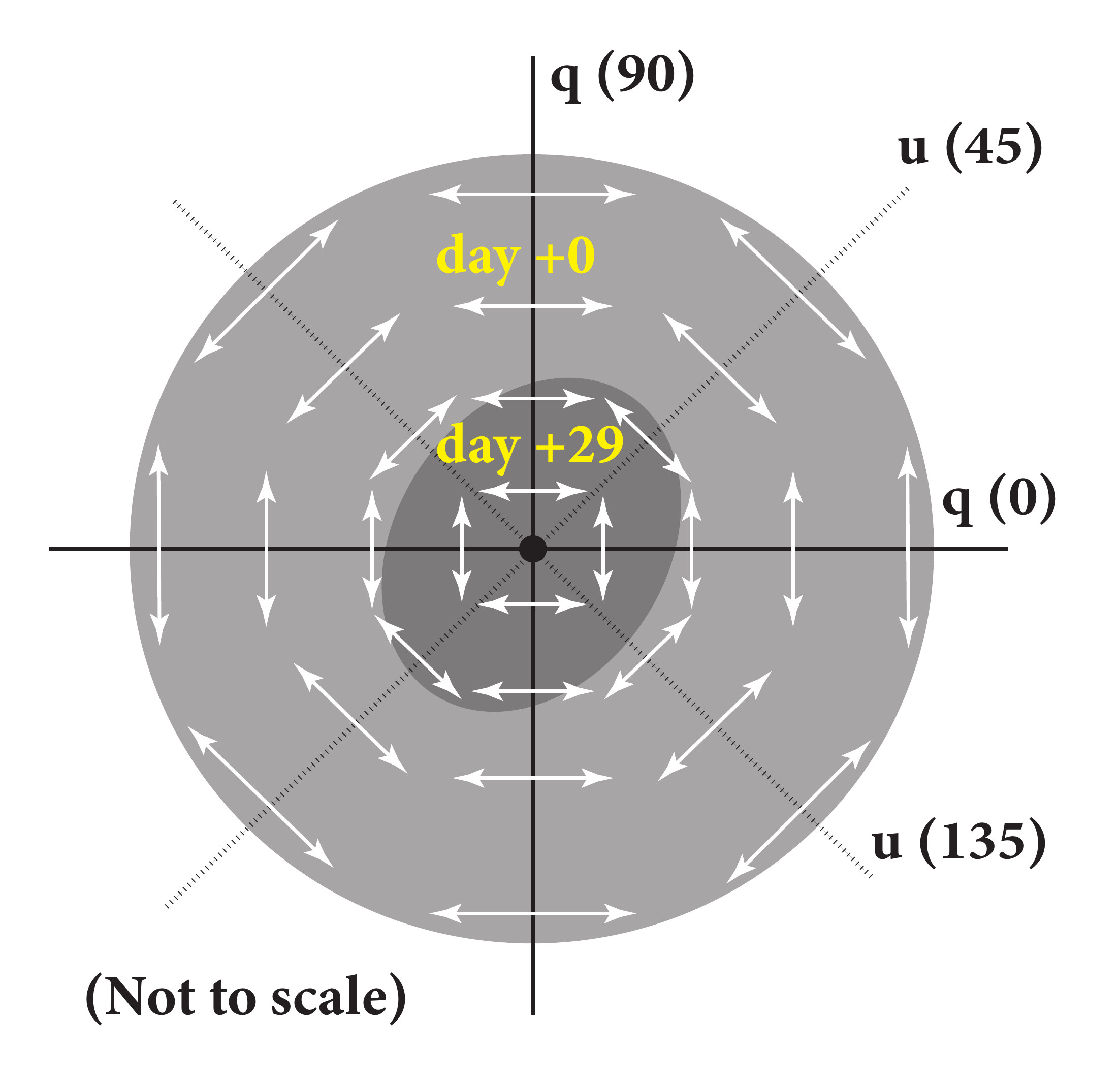}
	\caption{Schematic diagram illustrating the geometrical configuration of a sky-projected electron-scattering photosphere that is compatible with the observed polarimetry. The white arrows represent the relative size and direction of the local polarization. The outer circle shows the scattering photosphere on day $+0$, and the inner ellipse displays the photosphere a month later on day $+29$.  The inner ellipse has an aspect ratio of $e \approx 0.8$, which produces $\sim 1$ per cent continuum polarization in a scattering-dominated photosphere.}
	\label{fig:disc_geometry}
\end{figure}

Substructures in the photosphere can be probed by plotting the polarization in the Stokes $q-u$ plane. If the photosphere is axially symmetric, the data will lie along a straight line called the ``dominant axis'' \citep{Wang_etal_2003_01el, Maund_etal_2010}. In the $q-u$ plane, distance from the origin represents deviation from circular symmetry, whereas scatter around the dominant axis shows departure from axisymmetry (for e.g., clumps of material). Figures \ref{fig:adx_qu} and \ref{fig:aea_qu} show the polarization in the $q-u$ plane on days $+0$ and $+29$, respectively. The left panel displays polarization over the wavelength range 5400--8600\,\AA. Note that we ignore wavelengths below 5400\,\AA\ because the S/N on the blue end is poorer compared to the rest of the spectrum. The right-side panel also shows polarization but only for the H$\alpha$ feature. The dominant axes fitted to the data in each panel are indicated by the dashed black lines. The black star represents the error-weighted mean Stokes $q$ and $u$. At maximum light the data are distributed around the origin, indicating that the photosphere is highly spherical. On day $+29$, the data are farther from the origin, showing that an aspherical interior has begun emerging at this epoch. Additionally, as indicated by the increased value of chi-squared per degree of freedom ($\chi^{2}$/DoF), the scatter of the data relative to the dominant axis is also elevated on day $+29$ compared to day $+0$. The implication is that the aspherical photosphere is clumpier on day $+29$ than at maximum brightness. In the context of a TDE photosphere, the clumpiness of the material may refer to smaller-scale density variations that contribute different amounts of polarized flux. We sketch the implied geometry of the electron-scattering photospheres on days $+0$ and $+29$ in Figure \ref{fig:disc_geometry} based on the observed Stokes parameters, $PA$, and the relation between polarization and asphericity of scattering-dominated atmospheres \citep{Hoeflich_1991}.

\subsection{H$\alpha$ line polarization}

The H$\alpha$ emission line is depolarized compared to the continuum on day $+29$. This observation implies that the H$\alpha$ line-forming region is close to the electron-scattering surface, $r_{\text{th},\alpha} \approx r_{\text{s}}$. Our reasoning is as follows. If H$\alpha$ formed near the continuum thermalisation surface, the H$\alpha$ photons would be similarly polarized to the rest of the continuum as they diffuse through the electron-scattering-dominated region. Instead, if the H$\alpha$-forming region is closer to the electron-scattering surface, then the probability that H$\alpha$ photons will scatter is smaller owing to lower scattering optical depth, thus suppressing the polarization (see Figure 1 of \citealt{Hoeflich_1991}). Such depolarization is also seen for the He\,I emission line at 5875\,\AA.  It follows from this argument that the red shoulder of the H$\alpha$ profile should show a gradual increase in polarization --- the redder photons originate closer to the continuum thermalisation radius (as a result of lower line absorption opacity) --- until the polarization matches the continuum level at the end of the red shoulder. Hints of this are visible in the H$\alpha$ polarization profile of AT\,2019qiz on day $+29$, although the S/N is not sufficiently high to make a definitive claim. If this polarization feature is more securely realised in subsequent spectropolarimetric studies of TDEs, it would provide supporting evidence to the claim made by \citet{Roth_Kasen_2018} that electron scattering plays a significant role in spectral-line formation in TDEs.  

Alternatively, if the H$\alpha$ line actually forms outside the electron-scattering radius, spectropolarimetry can once again provide evidence. In that case, the H$\alpha$ emission line would be completely depolarized because flux from emission is intrinsically unpolarized.

\subsection{Comparison with other polarimetric studies of TDEs}

Owing to the low rate of TDEs ($\sim 10^{-6}$--$10^{-4}\,\text{yr}^{-1}$; \citealt{Holoien_etal_2016_tderates}) per galaxy, only a handful of polarimetric studies of TDEs currently exist. Nonetheless, these studies show a diverse range of TDE polarization. Here we summarise what is currently known about TDEs from polarimetric studies. 

The first reported spectropolarimetric measurement of a TDE was carried out by \citet{Holoein_etal_2020} for ASASSN-18pg, who found $\sim 1.5$ per cent continuum polarization on day $-10$. However, unlike for AT\,2019qiz, most of the polarization measured for ASASSN-18pg was attributable to ISP and not the TDE itself. Thus, our work is the first time intrinsic spectropolarimetric evolution of a TDE has been observed. 

Deep-infrared imaging polarimetry of the putative\footnote{The initial gamma-ray burst and X-ray properties were in conflict with a TDE interpretation, but late-time properties were more like a TDE \citep{Bloom_etal_2011}.} TDE {\it Swift} J164449.3+573451 showed $7.4 \pm 3.5$ per cent polarization \citep{Wiserma_etal_2012}. However, the authors were unsure about how much of the observed polarization was due to host-galaxy ISP in this highly-reddened event. {\it Swift} J2058+0516, which is comparatively less afflicted with reddening, exhibited polarization of $8.1 \pm 2.5$ per cent, with weak evidence for evolution over three epochs \citep{Wiserma_etal_2020}. Both {\it Swift} TDEs had relativistic jets in which high polarization may originate from the tail of nonthermal radiation (e.g., synchrotron) at optical/infrared wavelengths \citep{Lee_etal_2020_pol19dsg}. 

Optical imaging polarimetry of OGLE16aaa --- an unjetted thermal TDE --- showed polarization at the $1.81 \pm 0.42$ per cent level, but again the contribution of ISP is unknown \citep{Higgins_etal_2019}. Another thermal TDE, AT\,2019dsg, displayed high polarization of up to $\sim 9$ per cent in the optical $V$ band, which decreased to $\sim 3$ per cent a month later \citep{Lee_etal_2020_pol19dsg}. The authors attributed this high level of polarization to either an anisotropic accretion disc or contribution from relativistic jet emission. It is unclear how a TDE could achieve high polarization ($\sim 8$--9 per cent) by a purely electron-scattering process; \citet{Hoeflich_1991} showed that linear polarization maxes out at $\sim 4.5$ per cent in highly-oblate spheroids ($e < 0.2$; see Figure 4 of \citealt{Hoeflich_1991}). Thus, there is no clear explanation yet for the diversity of TDE polarization. Evidently, more polarimetric (spectroscopic or photometric) studies of TDEs are necessary to unify these observations.

\section{Conclusion}
\label{sec:conclusion}

We have presented 2 epochs of spectropolarimetry of the TDE AT\,2019qiz, demonstrating the first observed spectropolarimetric evolution of a TDE. The continuum polarization at maximum brightness was found to be consistent with 0 per cent, but increased to $\sim 1$ per cent a month later. These observations disfavour a naked accretion disc without significant outflow in AT\,2019qiz because high polarization is expected from an eccentric disc. Instead, our data favour the existence of a nearly spherical, optically thick, scattering-dominated gas layer. The apparent sphericity of the scattering photosphere explains the low polarization observed at peak brightness. We estimate the radius of the scattering photosphere to be $\sim 100\rm\,au$ at maximum brightness, which is much larger than the size of the complex, aspherical hydrodynamic structures of the bound gas.

As the outflow weakens and the photosphere recedes, the underlying aspherical interior begins to emerge, thus explaining the increased polarization on day $+29$. By plotting the polarization in the Stokes $q-u$ plane, we showed that on day $+29$ the electron-scattering layer is clumpy, which indicates the presence of smaller-scale density variations in regions contributing to the polarized flux. The depolarization of the H$\alpha$ emission line shows that the H$\alpha$ line-forming region is closer to the electron-scattering radius rather than to the continuum thermalisation radius.

This study demonstrates the immense power and potential of spectropolarimetry in delineating the geometry of a TDE, and thus providing a new perspective on the origin of the optical emission of TDEs. Indeed, well-sampled spectropolarimetric observations could help build a ``tomographic'' picture of a TDE as it evolves.  Primary challenges include the rarity of bright and nearby TDEs that will provide sufficiently high S/N to allow a feasible spectropolarimetric study. Additionally, no theoretical models for polarization of a TDE exist at the moment. Such models may enable stronger constraints on the TDE geometry as opposed to order-of-magnitude estimates. We hope that this work serves as a clarion call for further spectropolarimetric studies of TDEs.

\section*{Acknowledgements}

K.C.P. thanks Nina Pak for her help in creating Figure \ref{fig:disc_geometry}. We are grateful to Andrea Antoni, Ryan Chornock, and an anonymous referee for helpful discussions or suggestions. Ryan Foley recommended that we collect spectropolarimetric observations of AT\,2019qiz. A.V.F.'s group acknowledges generous support from the
Christopher R. Redlich Fund, the U.C. Berkeley Miller Institute for Basic Research in Science, Sunil Nagaraj, Landon Noll, Sandy Otellini, and many additional donors. W.L. was supported by the Lyman Spitzer,
Jr. Fellowship at Princeton University.
A major upgrade of the Kast spectrograph on the Shane 3\,m telescope          
at Lick Observatory, led by Brad Holden, was made possible through            
generous gifts from the Heising-Simons Foundation, William and Marina         
Kast, and the University of California Observatories. 
Research at Lick Observatory is partially supported by a generous gift from Google.
We appreciate the excellent assistance of the staff at Lick Observatory. 
NASA/IPAC Extragalactic Database (NED) 
is operated by the Jet Propulsion Laboratory, California Institute of Technology, 
under contract with NASA.

\section*{Data Availability}
The raw data used in this work may be shared upon request to Kishore C. Patra (kcpatra@berkeley.edu).



\bibliographystyle{mnras}
\bibliography{references} 







\bsp	
\label{lastpage}
\end{document}